\documentclass[a4paper,11pt]{article}
\usepackage{pos}
\usepackage{graphicx,epsfig,dcolumn,bm,epic,eepic,float}
\usepackage{amsmath}
\usepackage[T1]{fontenc}
\usepackage[utf8]{inputenc}
\usepackage{mathtools} 
\usepackage{pifont}
\usepackage{latexsym}
\usepackage{color}
\usepackage{cancel}
\usepackage{scrextend}
\usepackage{autobreak}
\usepackage{eucal}
\usepackage[bottom]{footmisc}
\usepackage{makeidx,shortvrb,latexsym}
\usepackage{epstopdf}
\usepackage{ragged2e}
\usepackage[numbers,sort&compress]{natbib}
\usepackage{epsfig, axodraw}
\usepackage{graphicx}    
\usepackage{graphics}
\usepackage{afterpage}
\usepackage{array}
\usepackage[flushleft]{threeparttable}
\usepackage{subcaption}
\usepackage{cancel}
\usepackage[mathlines]{lineno}
\usepackage{lettrine}
\usepackage{multirow}
\usepackage{longtable}
\usepackage{enumitem}

\renewcommand{\thefootnote}{\fnsymbol{footnote}}

\newcommand{\nc}{\newcommand}
\nc{\beq}{\begin{equation}}  \nc{\eeq}{\end{equation}}
\nc{\bea}{\begin{eqnarray}}  \nc{\eea}{\end{eqnarray}}
\nc{\baa}{\begin{array}}     \nc{\eaa}{\end{array}}
\pdfoutput=1

\DeclareMathAlphabet{\mathcal}{OMS}{cmsy}{m}{n}

\title{Mixed CP Violation and Natural Alignment in 2HDMs}

\author[a]{Neda Darvishi}
\author*[b,c]{Apostolos Pilaftsis} 

\affiliation[a]{Department of Physics, Royal Holloway, University of London,\\ Egham, Surrey, TW20 0EX, United Kingdom}
\affiliation[b]{Department of Physics and Astronomy, University of Manchester,\\ Manchester, M13 9PL, United Kingdom}
\affiliation[c]{PRISMA Cluster of Excellence \& Mainz Institute for Theoretical Physics,\\ Johannes Gutenberg University, 55099 Mainz, Germany}

\emailAdd{neda.darvishi@rhul.ac.uk}
\emailAdd{apostolos.pilaftsis@manchester.ac.uk}

\abstract{We present a new form of CP violation (CPV) that can be realised in Two-Higgs Doublet Models~(2HDMs) and was studied recently in~\cite{Darvishi:2023fjh}. By examining the vacuum manifold of a generic (convex) 2HDM potential, we identify scenarios that exhibit Mixed Spontaneous and Explicit CP Violation (MCPV), in which at least two {\em non}-degenerate CP-violating local minima coexist. We illustrate how this identification is achieved at the tree level by determining the magnitude and phase of a novel complex parameter, which we call~$r_{\rm CP}$. Since explicit CP Violation vanishes in 2HDMs where SM Higgs alignment is enforced through global continuous symmetries, we investigate how to maximise CPV in such scenarios by introducing soft or explicit breaking of the relevant symmetries. In doing so, we derive upper bounds on key CP-violating parameters that characterise misalignment with the SM, subject to constraints from the non-observation of the electron electric dipole moment. Finally, we delineate the region of the CP-violating parameter space in such constrained 2HDMs that can be further tested at the CERN Large Hadron Collider. }

\FullConference{}

\begin{document}
\maketitle 

\section{Introduction}

One of the simplest and most promising Standard Model (SM) extensions addressing these cosmological puzzles is the Two-Higgs-Doublet Model (2HDM), whose Higgs potential contains two SU(2)$_L$ scalar doublets, $\Phi_1$ and $\Phi_2$. The 2HDM can supply extra CP-violating sources~\cite{Lee:1973iz,Branco:1980sz,Branco:1985aq,Weinberg:1990me} crucial for electroweak baryogenesis~\cite{Cohen:1993nk}, and may also accommodate additional stable scalars acting as dark matter~\cite{Barbieri:2010mn}. Nevertheless, given the great phenomenological success of the SM in describing fundamental interactions of elementary particles at colliders, the coupling
strengths of the Higgs boson in the 2HDM, primarily to the electroweak~(EW) $W^\pm$ and $Z$ gauge bosons, must be very close to those predicted by the SM~\cite{ATLAS:2016neq,Palmer:2021gmo,ATLAS:2021upq}. Thus, any New Physics model must yield a 125~GeV scalar with SM-like properties in all its interactions. Such SM alignments arise either via fine-tuning~\cite{Chankowski:2000an,Gunion:2002zf,Ginzburg:2004vp,Carena:2013ooa} or naturally by imposing symmetries~\cite{Paschos:1976ay,Glashow:1976nt,Peccei:1977ur,
BhupalDev:2014bir,Darvishi:2019dbh}. In particular, the $\mathbb{Z}_2$ and U(1) symmetries in the 2HDM Yukawa sector~\cite{Glashow:1976nt,Peccei:1977ur} eliminate dangerous flavor-changing neutral currents at tree level.

In addition to Natural Flavour Conservation (NFC) in the Yukawa sector, Natural SM-Higgs Alignment (NHAL) in the gauge sector can be realised by imposing ${\rm SU(2)}_L$-preserving continuous symmetries~\cite{Pilaftsis:2016erj,BhupalDev:2014bir,Darvishi:2020teg,Darvishi:2019ltl,Darvishi:2021txa,Pilaftsis:2022euw}, without requiring heavy-state decoupling~\cite{Chankowski:2000an,Gunion:2002zf,Ginzburg:2004vp,Carena:2013ooa}. This symmetry-driven NHAL is independent of the ratio $\tan\beta = v_2/v_1$ of the $\Phi_{1,2}$ vacuum expectation values (VEVs), as well as the bilinear mass terms in the 2HDM potential. The only phenomenological requirement is a CP-even scalar of about 125~GeV, tied to spontaneous symmetry breaking in the so-called Higgs basis~\cite{GEORGI197995,PhysRevD.19.945,Lavoura:1994fv,Botella:1994cs}. In~\cite{BhupalDev:2014bir}, three continuous symmetry groups were identified that enforce NHAL at tree level: (i) the symplectic group ${\rm Sp}(4)$ (ii) ${\rm SU}(2)_{\rm HF}$ and (iii) ${\rm CP}\times {\rm SO}(2)_{\rm HF}$. Compared to an aligned scenario that relies heavily on fine-tuning in the Higgs basis, NHAL symmetries yield 2HDM potentials with all quartic couplings being real~\cite{BhupalDev:2014bir,Darvishi:2019ltl,Darvishi:2020teg}, thus eliminating explicit sources of CPV~\cite{Grzadkowski:2014ada}. Nevertheless, another possibility remains that we study here: SCPV in real Higgs potentials. Such SCPV scenarios require the third NHAL symmetry, $\text{CP}\times {\rm SO}(2)_{\rm HF}$, as well as two custodial symmetries, $\text{CP}\times\text{O}(4)$ and $\text{O}(2)\times \text{O}(3)$, with only soft CP-even mass terms.

In the literature~\cite{Lee:1973iz,Branco:1980sz,Branco:1985aq,Weinberg:1990me}, CPV is commonly attributed to two mechanisms: (a) explicit violation at the Lagrangian level, as in the SM with complex Yukawa couplings, or (b) vacuum-induced violation with at least two degenerate CP-violating vacua, separated by domain walls~\cite{Battye:2011jj, Battye:2020sxy}. These degenerate CPV vacua cannot be related by SM gauge transformations. Switching on small explicit CP-odd phases lifts the degeneracy but still yields two local CP-breaking minima, introducing a distinct Mixed CPV scenario where tunnelling between vacua can occur via a first-order EW phase transition. If the CP phases become sufficiently large and the tree-level 2HDM potential is convex, one obtains a unique global CPV vacuum (up to SM gauge transformations). A central result of our work in~\cite{Darvishi:2023fjh} was to identify a key complex parameter $r_{\rm CP}$ in the general 2HDM potential, whose magnitude $|r_{\rm CP}|$ and phase $\phi_{\rm CP}$ clearly classify three types of CPV: (i) Spontaneous, (ii) Explicit, and (iii) Mixed.

In this contributing note of the proceedings, we will also analyse another alternative for generating sizeable CPV in the scalar sector, whilst being close to NHAL. Specifically, we introduce small departures from NHAL through explicit CP-violating quartic couplings or through a complex bilinear mass term, $m^2_{12}\Phi^\dagger_1\Phi_2$, in the 2HDM potential. As a result, the larger the deviation from NHAL, the larger the CPV~is. Therefore, one objective of the note is to illustrate how to maximise~CPV, while maintaining agreement with the current LHC data, along with the strict upper limits on the electron EDM. Consequently, all misalignment directions that we classify in this second option will explicitly break CPV as well.

\setcounter{equation}{0}
\section{The CP-violating 2HDM}\label{sec:CPV2HDM}

The scalar potential $\mathcal{V}$ of the 2HDM which may be expressed in terms of the two Higgs doublets $\Phi_1$ and $\Phi_2$ as follows:
\begin{align}
   \label{eq:Vpot}
\mathcal{V} &= -\frac12 \bigg[ m_{11}^2 |\Phi_1|^2 +m_{22}^2 |\Phi_2|^2 + \Big(m_{12}^2 \Phi_1^\dagger \Phi_2+ {\rm H.c.} \Big)\bigg] + { \lambda_1} |\Phi_1|^4 +{ \lambda_2} |\Phi_2|^4 + \lambda_3 |\Phi_1|^2 |\Phi_2|^2
\nonumber	 \\
	 &+\lambda_4  |\Phi_1^\dagger \Phi_2|^2 + \bigg( \frac12 \lambda_5(\Phi_1^\dagger \Phi_2)^2+ \lambda_6 (\Phi_1^\dagger \Phi_2)|\Phi_1|^2+ \lambda_7 (\Phi_1^\dagger \Phi_2)|\Phi_2|^2+ {\rm H.c.} \bigg).
\end{align}
Note that the parameters, $m_{12}^2$ and $\lambda_{5,6,7}$, are complex, whereas the remaining parameters, $m^2_{11}$, $m^2_{22}$ and $\lambda_{1,2,3,4}$, are real. Our next step is to determine the ground state of the Higgs potential, and so study the topology of its vacuum manifold in its CP-odd phase or CP-odd scalar field direction. 

We start by considering the linear decompositions of the Higgs doublets,
\begin{equation}
  \label{Phi12}
\Phi_1\ =\ \left( \begin{array}{c}
\phi^+_1 \\ \frac{1}{\sqrt{2}}\, ( v_1\, +\, \phi_1\, +\, ia_1)
\end{array} \right)\, ,\qquad
\Phi_2\ =\ e^{i\xi}\, \left( \begin{array}{c}
\phi^+_2 \\  \frac{1}{\sqrt{2}}\, ( v_2 \, +\, \phi_2\, +\, ia_2 )
 \end{array} \right)\, ,
\end{equation}
where $v_1$ and $v_2$ are the moduli of the vacuum expectation values
(VEVs) of the Higgs doublets and $\xi$ is their relative phase. The key parameter $\tan\beta =  v_2/v_1$
and the SM VEV is $v  = \sqrt{v^2_1 + v^2_2}\,$.
 Here, we have adopted a weak basis in which VEVs $v_1,v_2$ and the quantum fluctuations $\phi_1,\phi_2$
hold the same phase. 

From the minimization condition the following general constraint on the CP phases of the 2HDM
potential can be deduced ~\cite{Darvishi:2023fjh}: 
 \begin{align}
{ 2\, |r_{\rm CP}| \sin(\xi+\varphi_{12})\: -\: \sin(2\xi+\varphi_5) }\: 
=\: 0 \,,
\label{MaxCP}
\end{align}
where the two phases are
\begin{align}
\varphi_5 \equiv \text{arg}(\lambda_5) \,=\, \tan^{-1}\left(\frac{\mathrm{Im}(\lambda_5)}{\mathrm{Re}(\lambda_5)}\right),\quad 
\varphi_{12} \equiv \tan^{-1}\left(\frac{\text{Im}\,\big(m^2_{12}  -\, 
                 \lambda_6\,c^2_\beta v^2\, -\, \lambda_7\,s^2_\beta v^2\big)}{\text{Re}\,\big(m^2_{12}  -\, 
                  \lambda_6\,c^2_\beta v^2\, -\, \lambda_7\,s^2_\beta v^2\big)}\right),
\end{align}
and
\begin{equation}
  \label{eq:rCP}
r_{\rm CP}\: \equiv\: \frac{m_{12}^2 -\,
                 \lambda_6\,c^2_\beta v^2\,  -\, \lambda_7\,s^2_\beta v^2}{\lambda_5s_{2\beta}\, v^2 }\ .  
\end{equation}
Observe that by writing $r_{\rm CP} = |r_{\rm CP}|\,e^{i\phi_{\rm CP}}$, 
we derive the key relation
$\phi_{\rm CP}\: =\: \varphi_{12}\, -\,
\varphi_5\: (\phi_{\rm CP} \in (-\pi\,, \pi]$). As we 
will see below, the values of $|r_{\rm CP}|$ and its phase $\phi_{\rm CP}$ play an instrumental role, as they allow us to identify the
different realisations of CPV in the 2HDM potential.
 
To simplify matters, we consider the weak-basis choice: $
\lambda_6\,=\, \lambda_7,\,\,{\rm Im} \lambda_5\, =\, 0\,,\,\, 
\lambda_5\, \ge\, 0. $
Thus one uniquely gets $\varphi_5 = 0$. This in turn implies that $\phi_{\rm CP} = \varphi_{12}$ and
\begin{equation}
  \label{eq:rCPabs}
|r_{\rm CP}|\, =\, \frac{\big|m_{12}^2\, -\,
                 \lambda_6\,v^2\big|}{\lambda_5 s_{2\beta}\,v^2 }\; .
\end{equation}
As shown in~~\cite{Maniatis:2011qu}, it is always possible to make a suitable weak-basis choice by exploiting the freedom of an 
SU(2) reparameterization of the two Higgs doublets, $\Phi_1$ and~$\Phi_2$, which modifies the 2HDM potential accordingly. This choice of weak basis uniquely fixes the scalar potential, by rendering the existence of possible accidental symmetries
manifest~\cite{Pilaftsis:1999qt,Battye:2011jj}. Consequently, in this same weak basis, one can unambiguously identify the three distinct sources of CP violation in a general 2HDM:
\vspace{-0.3cm}
\begin{enumerate}[itemsep=-0.05cm]
\item[{\bf (i)}] {\bf Spontaneous CP Violation (SCPV)}~\cite{Lee:1973iz,Branco:1980sz,Branco:1985aq}, where only $\xi\neq 0$ and all 2HDM-potential parameters remain real, i.e.~${\rm Im}(\lambda_6) = {\rm Im}(\lambda_7)=0$ and ${\rm Im}(m^2_{12}) =0$, so that $\phi_{\rm CP} = 0$ or $\pi$. From the bound in~\eqref{MaxCP}, this implies $0<|r_{\rm CP}| < 1$. In this case, the vacuum manifold $\mathcal{M} \;=\;\bigl\{\,(v_1,\,v_2 \,e^{i\xi}),\,(v_1,\,v_2 \,e^{-\,i\xi})\bigr\},$ consists of two disconnected degenerate neutral vacua,
which cannot be related by SM gauge transformations and hence may induce the formation of domain walls~\cite{Battye:2011jj,Chen:2020soj,Battye:2020sxy}.

\item[{\bf (ii)}] {\bf Explicit CP Violation (ECPV)}~\cite{Weinberg:1990me,Pilaftsis:1999qt}, i.e.~${\rm Im}(\lambda_6) = {\rm Im}(\lambda_7)\neq 0$ and ${\rm Im}(m^2_{12}) \neq 0$. In the ``pure'' ECPV limit, $|r_{\rm CP}| > 1$, and only a single global minimum exists if the potential is bounded from below. According to the CP-odd tadpole conditions, the phase $\xi$ here may also take CP-conserving values. According to the CP-odd tadpole conditions, the phase $\xi$ here may also take CP-conserving values (0 or~$\pi$) when ${\rm Im}(m^2_{12}) = {\rm Im}(\lambda_6) v^2$.

\item[{\bf (iii)}] {\bf Mixed Spontaneous and Explicit CP Violation (MCPV)}~\cite{Darvishi:2023fjh}, with three non-zero physical CP phases:
$$
\xi \;\neq\; 0,\quad
{\rm Im}(\lambda_6) \;=\; {\rm Im}(\lambda_7)\;\neq\; 0, \quad
{\rm Im} (m^2_{12}) \;\neq\; 0.
$$
In MCPV, the 2HDM potential admits two local CP-violating minima
~$\mathcal{V}(v_1,v_2 
e^{i\xi_1}) < \mathcal{V}(v_1,v_2 e^{i\xi_2})$,
where the lower-energy minimum corresponds to $\xi=\xi_1$ and the higher-energy one to $\xi=\xi_2$. 
\end{enumerate}

\begin{figure}[t]
\centering
\includegraphics[width=0.45\textwidth]{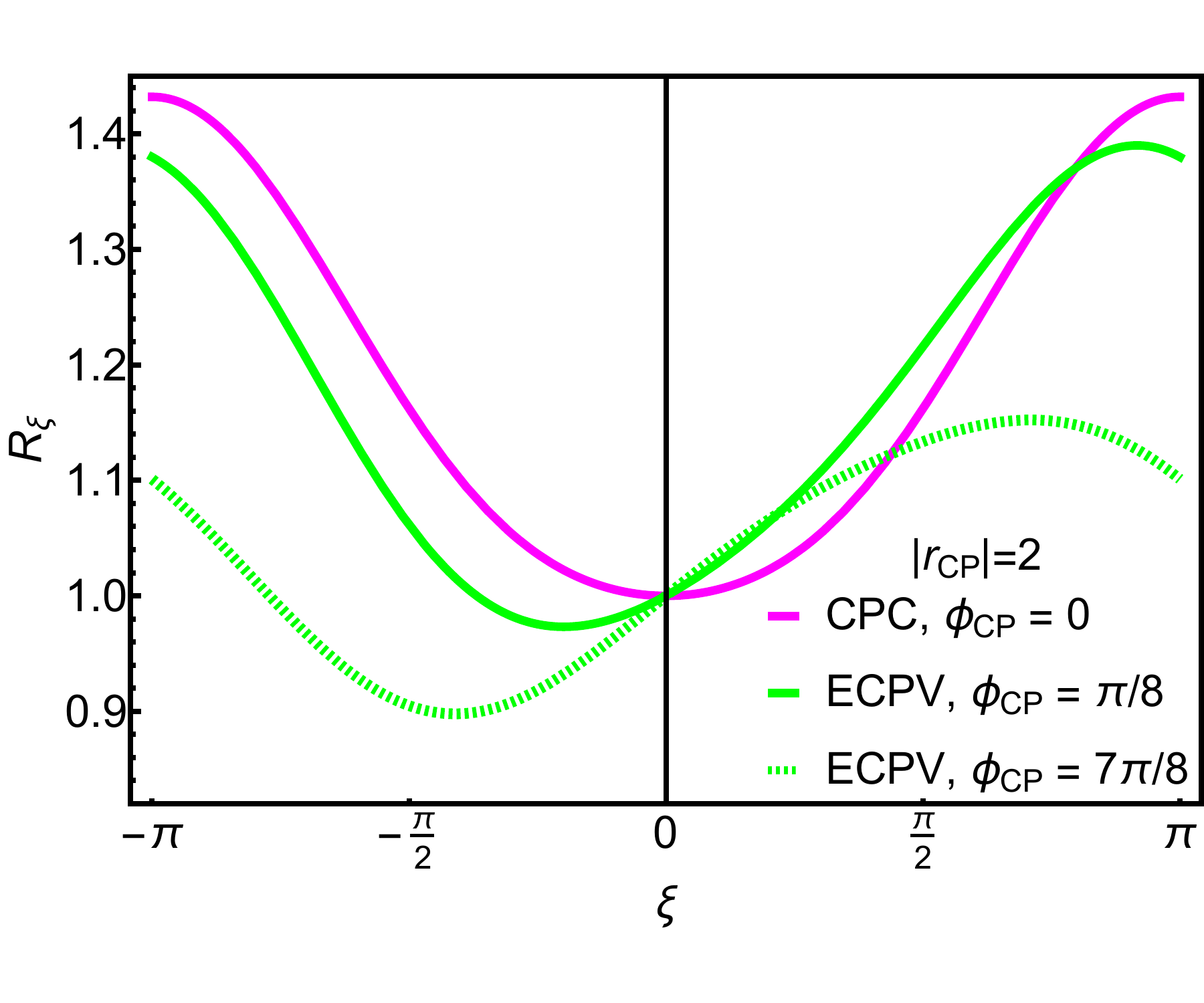}
\includegraphics[width=0.45\textwidth]{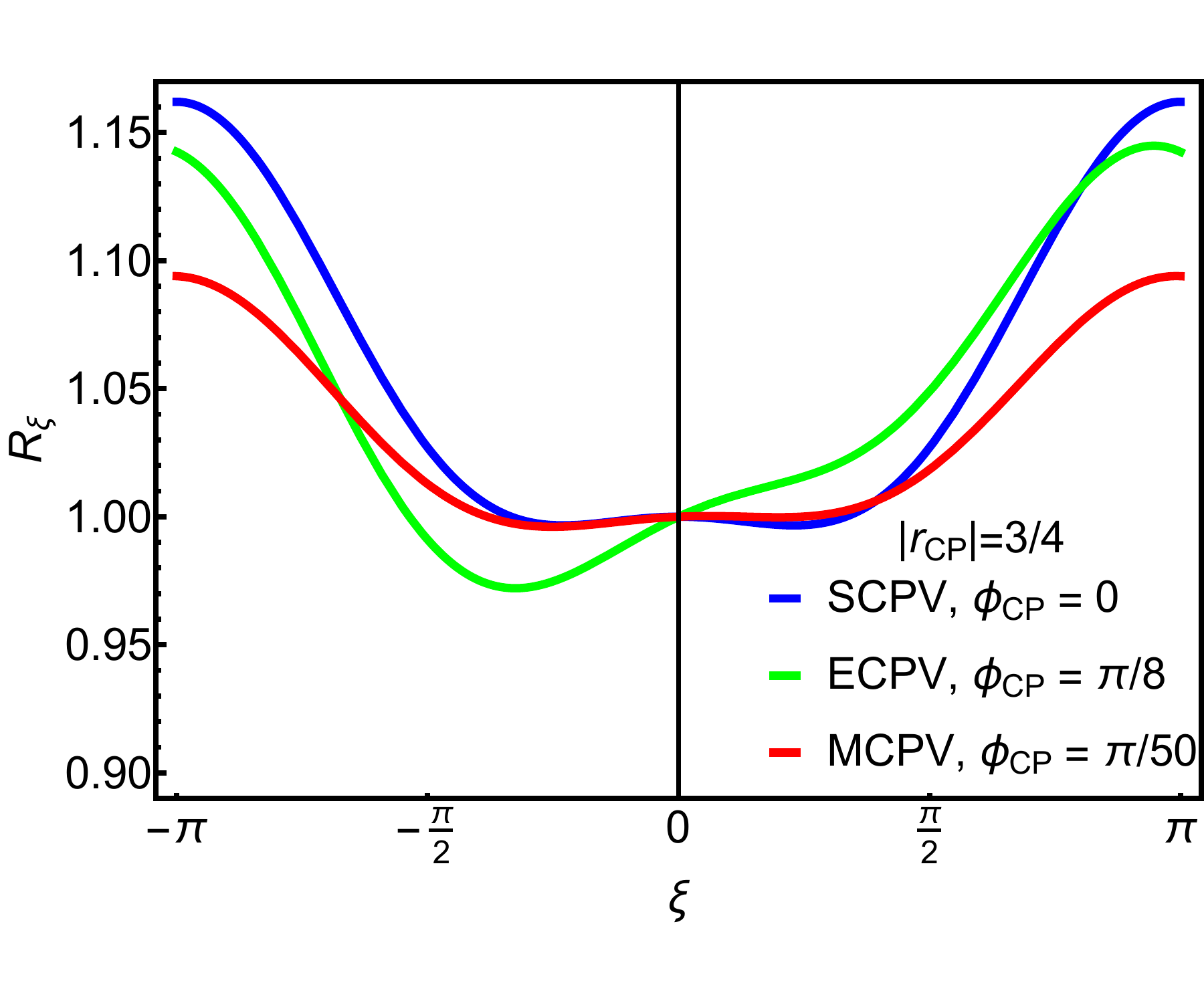}\vspace{-0.25in}
\includegraphics[width=0.45\textwidth]{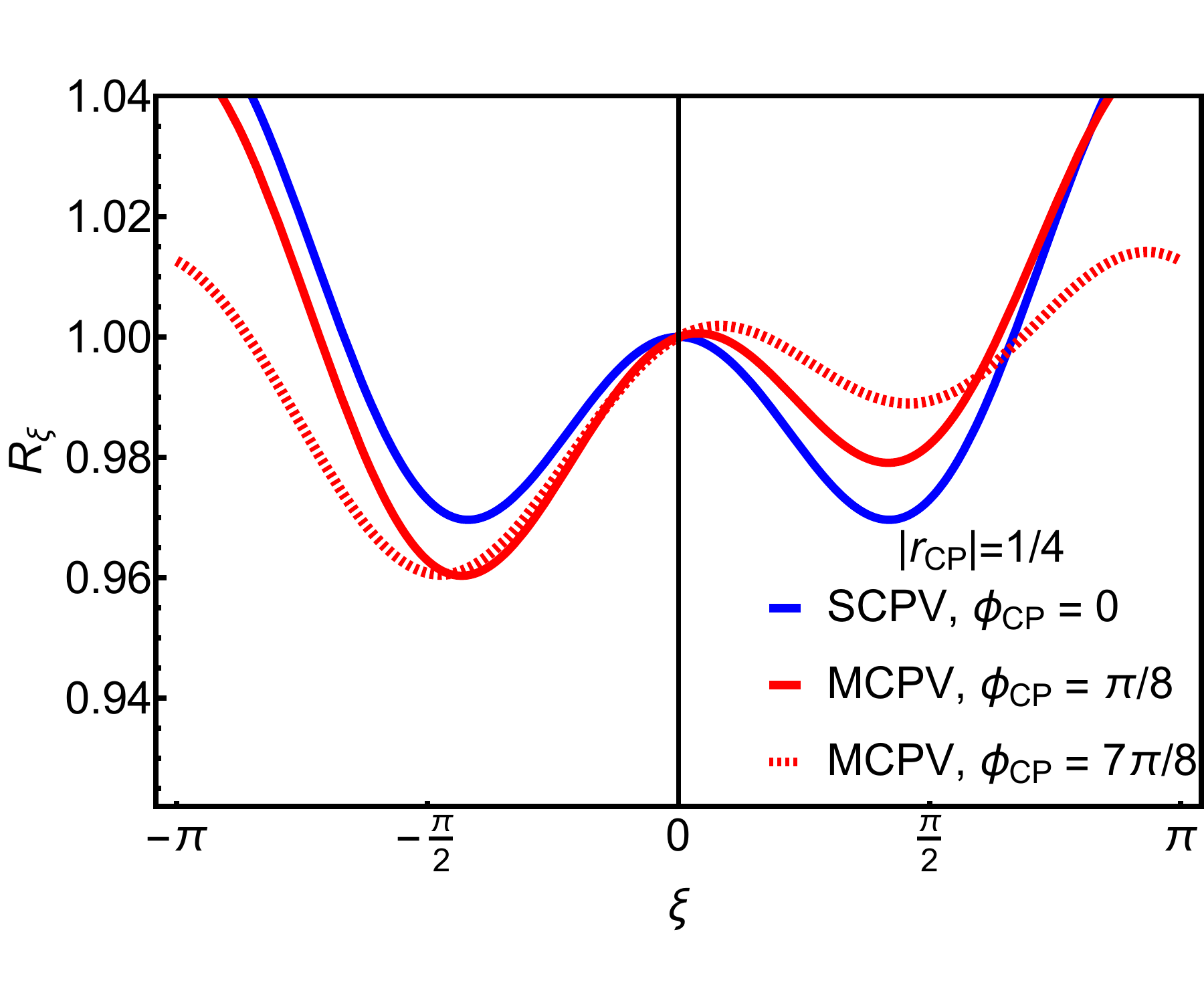}
\includegraphics[width=0.45\textwidth]{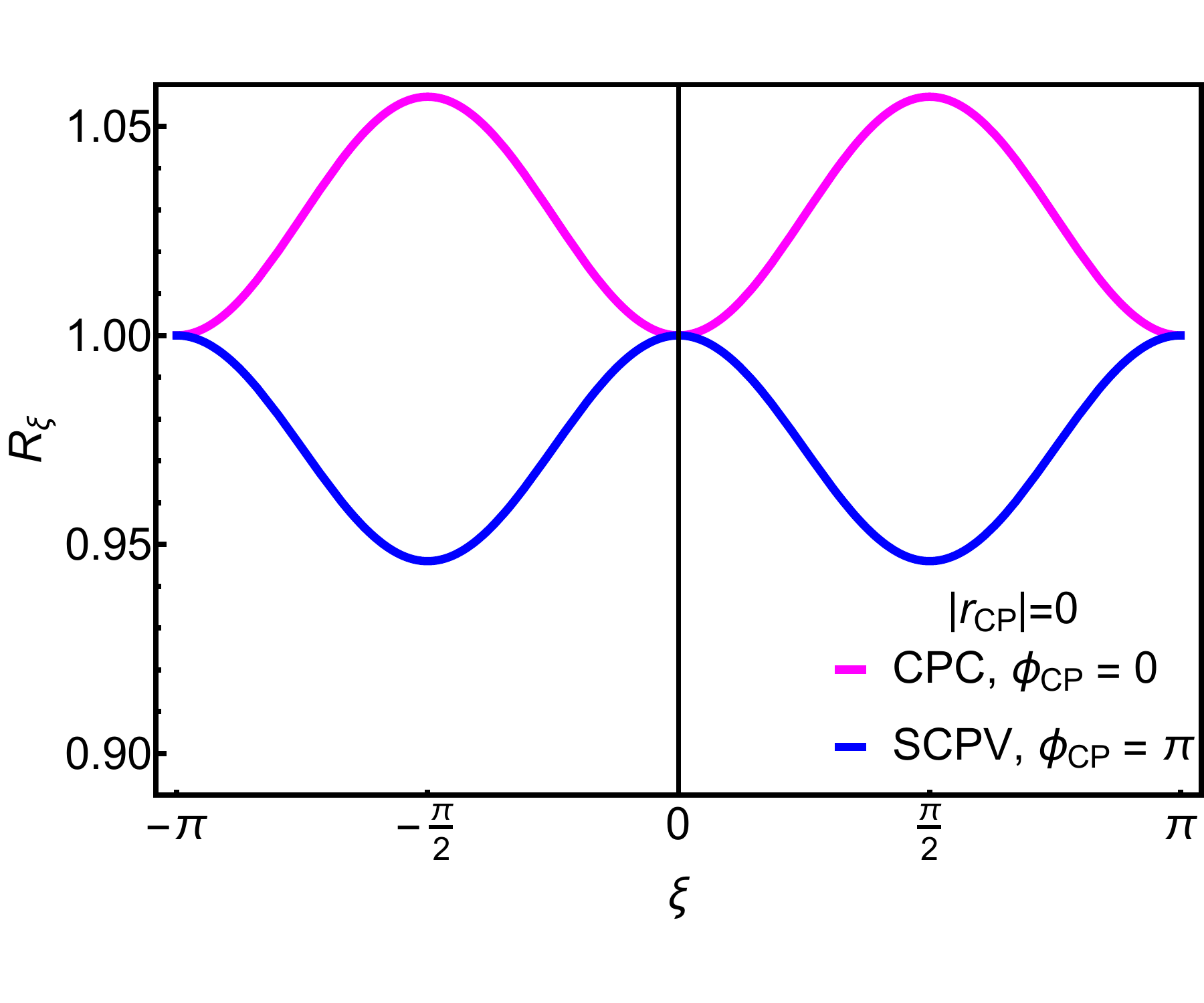}
\caption{\textit{The ratio $R_\xi\equiv \mathcal{V}(v_1,v_2 e^{i\xi})/|\mathcal{V}(v_1,v_2)|$ versus the phase $\xi$ is shown for selective values of $|r_{\rm CP}|$ and $\phi_{\rm CP} = \varphi_{12}$, and $\tan\beta = 2$. The distinct vacuum topologies of the 2HDM potential for CPC (magenta), ECPV (green), SCPV (blue), and MCPV (red) are illustrated.}}
\label{fig:Rxi}
\end{figure}

To illustrate the above three realisations of CPV that can take place in the general 2HDM potential~\eqref{eq:Vpot}, below, we introduce a convenient quantity, 
\begin{equation}
\label{eq:Rxi}
R_\xi \;\equiv\; {\mathcal{V}_\xi \over \,|\mathcal{V}_{\xi=0}|\,}\;=\;\mathrm{sgn}\,D\;+\;{N \over \,|D|\,},
\end{equation}
where $\mathrm{sgn}(x)=+1(-1)$ if $x$ is positive (negative), and
\begin{equation}
\mathcal{V}_\xi \;\equiv\; \mathcal{V}(v_1,v_2\,e^{i\xi}) 
\;=\; {v^2\over \,32\,}\,\Big[D + N(\xi)\Big],
\qquad
\mathcal{V}_{\xi=0} \;=\; {v^2 \over \,32\,}\,D,
\end{equation}
with $N(\xi)$ capturing the nontrivial $\xi$-dependence and $N(0)=0$. 

Although the CP-even tadpole conditions strictly determine $v_{1,2}$ only at the global minimum $\xi=\xi_{\rm min}$, here we fix $\xi=0$ as a reference value and eliminate $m_{11}^2$ and $m_{22}^2$ by minimization in favour of other potential parameters. In this way, one obtains a clear depiction of the vacuum topology along the $\xi$ direction, revealing how the 2HDM potential may develop CP-violating minima. In the weak basis we have
\begin{align}
\label{eq:Vxi}
\mathcal{V}_\xi\ =&\ { v^2\over 32}\bigg\{v^2 \big( 6 \lambda_1 + \lambda_3 + \lambda_4\big) -
    4 \big(m_{11}^2 + m_{22}^2\big)- 4 \big( m_{11}^2 - m_{22}^2\big) c_{2\beta}  +  
    v^2 \big(2 \lambda_1 - \lambda_3 - \lambda_4\big) c_{4\beta}
     \nonumber   \\ -&\
2 s_{2\beta} \Big[ 4 \Big({\rm Re}(m_{12}^2) - {\rm Re}(\lambda_6) v^2\Big) \cos\xi   - \lambda_5 s_{2\beta} v^2 \cos 2 \xi  - 4 \Big({\rm Im}(m_{12}^2) -{\rm Im}(\lambda_6) v^2 \Big) \sin\xi\Big]\bigg\}\;.
\end{align}
With the help of the analytic expression in~\eqref{eq:Vxi}, the quantities $N$ and $D$ in~\eqref{eq:Rxi} are 
found to be
\begin{align}
  \label{eq:Nparam}
N\, =& 
\, 4\,|\lambda_5| s_{2\beta}^2 v^2\,\Big[\, 2\, |r_{\rm CP}| \Big(\cos\varphi_{12} - \cos(\xi +\varphi_{12}) \Big)\: -\: \text{sgn}(\lambda_5)\,\sin^2\xi \,\Big],
\\
  \label{eq:Dparam}
D\,=&\ -v^2 \Big[6 \lambda_1+\lambda_3+\lambda_4  +c_{4 \beta} \Big(2 \lambda_1-\lambda_3-\lambda_4
-\lambda_5\Big) + \lambda_5 + 8 s_{2 \beta} {\rm Re}(\lambda_6)\Big]\;. 
\end{align}
Note that the overall minus sign in the quantity $D$ in~\eqref{eq:Dparam} that emerged after eliminating the
mass parameters $m^2_{11}$ and $m^2_{22}$. In addition, we set $\lambda_1 =\lambda_2$, to show the profile of the scalar potential as a function of the CP-odd phase $\xi$ in our examples.

In Figure~\ref{fig:Rxi}, we show the different vacuum topologies that determine the shape of the normalised potential $R_\xi$ [cf.~\eqref{eq:Rxi}] along the phase direction $\xi \in (-\pi,\pi]$, for illustrative choices of $|r_{\rm CP}|$ and $\phi_{\rm CP}$. Moreover, we keep the value of $\tan\beta$ fixed to a specific value, i.e.~${\tan\beta = 2}$ or~$s_{2\beta} = 4/5$. More explicitly, depending on the value of $|r_{\rm CP}|$, the following CPV patterns are observed~\cite{Darvishi:2023fjh}:
\vspace{-0.3cm}
 \begin{enumerate}[itemsep=-0.05cm]
\item[{\bf (a)}] {\boldmath $|r_{\rm CP}| >1$}. Then, the 2HDM potential exhibits either CP Conservation (CPC) or ECPV. This means that in addition to CPC if $\phi_{\rm CP} = 0$ or $\pi$, only ECPV is possible with a global CPV minimum and a local CPV maximum when $\phi_{\rm CP} \neq 0,\, \pi$, i.e.~the transcendental equation~\eqref{MaxCP} has 2 roots. These features are illustrated in Figure~\ref{fig:Rxi}~(upper left panel).
\item[{\bf (b)}] {\boldmath $1/2 \leq |r_{\rm CP}| < 1$}. Then, depending on $\phi_{\rm CP}$, the 2HDM potential will always violate CP in any of the three different forms, SCPV (when $\phi_{\rm CP}=0,\,\pi$), MCPV for small non-zero $\phi_{\rm CP}$ values, and ECPV otherwise. These possibilities were presented in Figure~\ref{fig:Rxi}~(upper right panel). We note that for $|r_{\rm CP}| = 1/2$, the critical $\phi_{\rm CP}$ value for the transition from MCPV to ECPV is $\phi_{\rm CP} = \pm \pi/4$. From~\eqref{MaxCP}, we find for the critical point $\phi_{\rm CP}  = \varphi_{12} = \pm\pi/4$ three roots in the principal branch of~$\xi \in (-\pi, \pi]$: $\xi_1 = \pm\pi/4$ (saddle point), $\xi_2 = \mp5\pi/12$ (local minimum) and $\xi_3 = \pm 11\pi/12$ (local maximum).
\item[{\bf (c)}] {\boldmath $0< |r_{\rm CP}| <1/2$}. In this case, the 2HDM potential will always violate CP either spontaneously, or
in a MCPV manner if $\phi_{\rm CP} \neq 0,\, \pi$.
Moreover, the transcendental equation~\eqref{MaxCP} will have four roots in the primary interval $\xi \in (-\pi,\pi]$, see Figure~\ref{fig:Rxi}~(lower left panel).
\item[{\bf (d)}] {\boldmath $r_{\rm CP} = 0$}. The extremal CP-odd constraint~\eqref{MaxCP} has four solutions: $\xi = 0,\, \pm\pi/2,\, \pi$, if $\lambda_5$ is non-zero and its phase $\varphi_5 = 0$ or $\pi$. For $\varphi_5 = \pi$, the extrema would correspond to two CPC minima at $\xi = 0,\, \pi$, as well as to two local CPV maxima at $\xi=\pm\pi/2$, as shown in Figure~\ref{fig:Rxi}~(lower right panel). The swap~$\varphi_5 = 0$ interchanges the roles of the minima and maxima, allowing for SCPV.  In both cases, we must have $M^2_{H^\pm}/v^2 > 0$ or $|\lambda_5| > \lambda_4$, if $\lambda_4 > 0$. Also, the breaking of the electroweak symmetry can lead to domain walls. In~addition, if ${\rm Im}\,\lambda_6 \neq 0$, the theory will violate CP explicitly, e.g.~through non-zero scalar-pseudoscalar mass terms~\cite{Pilaftsis:1999qt}. However, the latter source of CPV will not affect the profile of the tree-level 2HDM potential. Its effect will only show up beyond the Born approximation by lifting any degeneracy between minima and inducing radiative MCPV. 

\end{enumerate}

Finally, it is worth stressing that, in all the scenarios listed above, only two CPV phases can be at most independent of each other in the general 2HDM. Any third phase would be fixed by the CP-odd tadpole condition.

\setcounter{equation}{0}
\section{Continuous Symmetries and Natural Alignment}
\label{sec:CSNAL}

In this section, we first briefly review the 2HDM potentials on which NHAL is enforced by continuous symmetries. We then identify candidate scenarios that enable sizeable CPV while maintaining agreement with alignment constraints on the SM-like Higgs boson couplings to the EW $W^\pm$ and $Z$ bosons. A suitable framework to address this topic is the covariant bilinear field formalism introduced in~\cite{Maniatis:2006fs,Nishi:2006tg,Ivanov:2006yq,Battye:2011jj,Pilaftsis:2011ed,Pilaftsis:2024uub}. 

To start with, we introduce an $8$-dimensional complex $\bm{\Phi}$-multiplet that represents a vector in the SU(2)$_L\times$Sp(4) field space. The $\bm{\Phi}$-multiplet consists of the scalar iso-doublets~$\Phi_i$ (with $i=1,2$) and their $\mathrm{U(1)}_Y$ hypercharge-conjugate counterparts, $\widetilde{\Phi}_i = i \sigma^{2} \Phi_i^{*}$, i.e. 
\bea
   \label{eq:8DPhi}
\bm{\Phi}^{\sf T} = \begin{pmatrix}
\,\Phi_1, \,\Phi_2, \,\widetilde{\Phi}_1, 
\,\widetilde{\Phi}_2\:\end{pmatrix}^{\sf T}\,, 
\eea
 with $\sigma^{1,2,3}$ being the Pauli matrices. The ${\bm{\Phi}}$-multiplet has three essential properties. It transforms covariantly under separate $\mathrm{SU(2)}_L$ and Sp(4) transformations and obeys a Majorana-type constraint~\cite{Battye:2011jj}: 
\begin{equation}
  \label{eq:su2sp4}
\text{(i)}~{\bm{\Phi}'}\, =\,  U_L \, {\bm \Phi}\;,\qquad
\text{(ii)}~{\bm{\Phi}'}\, =\,  U\, {\bm \Phi}\;,\qquad
\text{(iii)}~{\bm{\Phi}}\, =\, C\, {{\bm \Phi}^*}\;, 
\end{equation}
with $U_L\in \mathrm{SU(2)}_L$ and $U\in \text{Sp}(4)$. Moreover,  $C=\sigma^2 \otimes {\bf{1}}_2 \otimes \sigma^2$ is the charge conjugation matrix, which also equips the Sp(4) space with a metric: $U C U^{\sf T} = C$, with~${C=C^{-1}=C^*}$.

Accordingly, the bilinear field vector may be defined as $
R^A \,\equiv\, {\bm{\Phi}}^{\dagger} \, {\Sigma}^{A}\, {\bm{\Phi}} \,(A=0, \,1,\,\cdots,\,n(2n-1)-1)
$ ~\cite{Ivanov:2006yq,Pilaftsis:2011ed,Battye:2011jj,Darvishi:2020teg}.
The ${\Sigma}^{A}$ matrices
may be expressed in terms of double tensor products as
\bea
{\Sigma}^{A} \, = \, \big(\sigma^0 \otimes t^a_S \otimes \sigma^0, \,\, \sigma^i \otimes t^b_A \otimes \sigma^0 \big),
\eea
with $t^a_S\,\;(t^b_A)\,\in \mathrm{SU}(n)$ as the symmetric (anti-symmetric) generators. 
Therefore, by virtue of $R^A$, the potential $\mathcal{V}$ can now be written down in the following quadratic form:
\bea
\mathcal{V} \,=\, -{1\over 2} \,M_A \,  R^A \,+\, {1 \over 4} \, L_{AB} \,  R^{A} \,  R^{B},
\label{VB}
\eea
the vector $M_{A}$ and the tensor $L_{{A}{B}}$ contain the mass parameters and quartic couplings of the 2HDM potential.

Thus far, several studies have established~\cite{Pilaftsis:2016erj,Darvishi:2019dbh,Darvishi:2020teg,Birch-Sykes:2020btk} that the potential $\mathcal{V}$ of the 2HDM contains 13 ${\mathrm{SU(2)}}_L$-preserving accidental symmetries as subgroups of
the maximal symmetry $\text{SU}(2)_L\otimes \text{SO}(5)$, of which 6 symmetries are $\text{U(1)}_Y$ invariant~\cite{Ivanov:2006yq}.
The symmetry group, $\mathrm{SU}(2)_L\otimes \mathrm{SO}(5)$, which
acts on the 5D bilinear field sub-space $R^I$ (with $I=1,2,3,4,5$), is isomorphic to $\mathrm{SU}(2)_L \otimes \text{Sp}(4)/Z_2$ in the original field space. It plays an instrumental role in classifying accidental symmetries that may occur in the scalar potentials of 2HDM and 2HDM-Effective
Field Theories (2HDMEFT) with higher-order
operators. These classifications were done for the
2HDM in~\cite{Battye:2011jj,Pilaftsis:2011ed,Darvishi:2019dbh} and for the 2HDMEFT framework in~\cite{Birch-Sykes:2020btk}.

\begin{table}[ht]
\small
\begin{center}
\begin{tabular}{ |c|l|l|l| } 
 \hline
No. & Generators & Continuous Syms & Parameters\\ \hline 
\hline
$8$  & $T^{0-9}$ & SO(5) & $\ m_{11}^2= m_{22}^2,  \, \lambda_1=\lambda_2=\lambda_3/2.$ \\
\hline 
$7\,\,$ & $T^{0,2,5,7,8,9}$ & CP1$\times$O(4)&$\ m_{11}^2= m_{22}^2,  \, \lambda_1=\lambda_2 = \lambda_3/2\,, \lambda_4= -{\rm Re}(\lambda_5).$ \\
$7'\,$  & $T^{0,1,4,6,8,9}$ & CP2$\times$O(4)$'$&$\ m_{11}^2= m_{22}^2,  \, \lambda_1=\lambda_2 = \lambda_3/2,\, \lambda_4={\rm Re}(\lambda_5).$ \\
$7''$  & $T^{0,3,4,5,6,7}$ & Z$_2\times$O(4)$''$&$\ m_{11}^2= m_{22}^2,  \, \lambda_1=\lambda_2,\, \lambda_3.$ \\
\hline
$6\,\,$  & $T^{0,2,5,7,8,9}$ & SO(4) &$m_{11}^2=m_{22}^2,\,  {\rm Im}( m_{12}^{2}),\, \lambda_1= \lambda_2=\lambda_{3}/2,\,\lambda_4= -{\rm Re}(\lambda_5),$
\\
& & & $\,{\rm Im}(\lambda_6)={\rm Im}( \lambda_7).$\\
$6'\,$  & $T^{0,1,4,6,8,9}$ &  SO(4)$'$ &$m_{11}^2=m_{22}^2,\,  {\rm Re}( m_{12}^2),  \, \lambda_1= \lambda_2=\lambda_{3}/2,\,\lambda_4= {\rm Re}\,(\lambda_5),$\, 
\\
& & & $\, {\rm Re}(\lambda_6)={\rm Re}(\lambda_7).$\\
$6''$  & $T^{0,3,4,5,6,7}$ & SO(4)$''$&$\ m_{11}^2,\, m_{22}^2,  \, \lambda_1,\,\lambda_2,\,\lambda_{3}.$ \\
\hline
$5\,\,$ & $T^{0,2,4,6}$ & O(2)$\times$ O(3)&$\ m_{11}^2=m_{22}^2, \, \lambda_1=\lambda_2=\lambda_{345}/2,\,\lambda_4={\rm Re}(\lambda_5).$\\
$5'\,$ & $T^{0,1,5,7}$  &   O(2)$'\,\times$ O(3)&$\ m_{11}^2=m_{22}^2, \, \lambda_1= \lambda_2= \bar{\lambda}_{345}/2,\,\lambda_4=-{\rm Re}(\lambda_5).$\\
$5''$ & $T^{0,3,8,9}$ &    O(2)$''$ $\times$ O(3) & $\ m_{11}^2=m_{22}^2, \, \lambda_1=\lambda_2=\lambda_{3}/2,\,\lambda_4.$\\
\hline
$4$  & $T^{0-3}$ &  O(3)$\times$ O(2)$_Y$ &$\ m_{11}^2=m_{22}^2, \, \lambda_1=\lambda_2= \lambda_{34}/2.$\\
\hline
$3\,\,$ & $T^{0,5,7}$ &  SO(3)&$m_{11}^2,\, m_{22}^2,\,  {\rm Im}( m_{12}^2),  \, \lambda_1,\, \lambda_2,\, \lambda_{3},\,\lambda_4= -{\rm Re}(\lambda_5), \, {\rm Im}(  \lambda_6),\, {\rm Im}(\lambda_7).$
\\
$3'\,$ & $T^{0,4,6}$ &  SO(3)$'$&$m_{11}^2,\, m_{22}^2,\, {\rm Re}( m_{12}^2),  \, \lambda_1,\, \lambda_2,\, \lambda_{3},\,\lambda_4={\rm Re}(\lambda_5),\, {\rm Re}(\lambda_6),\,{\rm Re}(\lambda_7).$
\\
$3''$ & $T^{0,8,9}$ & SO(3)$''$&$\ m_{11}^2=m_{22}^2,\,  m_{12}^2,  \, \lambda_1=\lambda_2=\lambda_{3}/2,\,\lambda_4,\, \lambda_5,\,\lambda_6=\lambda_7.$
\\
\hline
$2\,$ & $T^{0,2}$ & CP1$\times$O(2)$\times$O(2)$_Y$ &$ m_{11}^2=m_{22}^2, \,\lambda_1=\lambda_2=\lambda_{345}/2.$ \\
$2'\,\,$ & $T^{0,1}$ & CP2$\times$O(2)$'\times$O(2)$_Y$ &$ m_{11}^2=m_{22}^2,\,  \lambda_1=\lambda_2=\bar{\lambda}_{345}/2.$ \\
$2''$ & $T^{0,3}$ & S$_2\times$O(2)$''\times$O(2)$_Y$ &$ m_{11}^2=m_{22}^2,\,  \lambda_1=\lambda_2,\, \lambda_3,\ \lambda_4.$ \\
\hline
$1\,\,$ & $T^{0,2}$ & O(2)$\times$O(2)$_Y$&$ m_{11}^2=m_{22}^2, \, {\rm Im}(m_{12}^{2}),  \,\lambda_1=\lambda_2=\lambda_{345}/2,\,{\rm Im}(\lambda_{6})={\rm Im}(\lambda_{7}).$
\\
$1'\,$ & $T^{0,1}$ & O(2)$'\times$O(2)$_Y$ &  $ m_{11}^2=m_{22}^2, {\rm Re}(m_{12}^{2}),\,  \lambda_1=\lambda_2=\bar{\lambda}_{345}/2,\,{\rm Re}(\lambda_{6}) ={\rm Re}(\lambda_{7}).$
\\
$1''$ & $T^{0,3}$  &  O(2$)''\times$O(2)$_Y$ & $\ m_{11}^2,\,m_{22}^2,  \, \lambda_1,\,\lambda_2,\, \lambda_{3},\,\lambda_4.$\\
\hline
\end{tabular}
\end{center}
\caption{\textit{NHAL and next-to-NHAL symmetries of the 2HDM along with the various SO(5) generators as defined in~\cite{Pilaftsis:2011ed} that reinforce these symmetries. The relations between non-zero parameters associated with these symmetries are given, where we used the abbreviations: $\lambda_{34} \equiv \lambda_3 + \lambda_4$, $\lambda_{345} \equiv \lambda_3 + \lambda_4 + {\rm Re}(\lambda_5)$
and $\bar{\lambda}_{345} \equiv \lambda_3 + \lambda_4 -{\rm Re}(\lambda_5)$. Note that all product groups are subsets of SO(5), so their overall determinant must be evaluated to~1.}}
\label{tab:1}
\end{table}

In Table~\ref{tab:1}, we list all continuous symmetries of the 2HDM potential~\cite{Battye:2011jj,Pilaftsis:2011ed,Darvishi:2019dbh}, where the relevant choices of $\mathrm{SO(5)}$ generators are given according to the conventions of~\cite{Pilaftsis:2011ed}. This table also provides the relations among the non-zero mass and quartic-coupling parameters implied by these symmetries. Notably, some of these continuous symmetries can accommodate NHAL in a way that is largely independent of both the size and structure of the soft-symmetry-breaking mass parameters that may be introduced, as well as of any specific value of $\tan\beta$~\footnote{Alternatively, SM alignment can be achieved by imposing exact discrete symmetries~\cite{Darvishi:2020teg}, yielding  
$$
\tan\beta=1\:,\, \lambda_1\ =\ \lambda_2, \, \ \lambda_{3}, \, \ \lambda_{4}, \, \ \lambda_{5}, \, 
\lambda_6\ =\ \lambda_7\ .
$$
However, such parameter relations mainly lead to an inert Type-I 2HDM in the Higgs basis, for which~NHAL becomes automatic.}. Thus, NHAL is obtained under the following condition ~\cite{BhupalDev:2014bir,Darvishi:2020teg},
\begin{equation}
  \label{eq:NAcond}
\lambda_1\ =\ \lambda_2\ =\ \lambda_{345}/2, \qquad 
\lambda_6\ =\ \lambda_7\ =\ 0\, . 
\end{equation}
with $\lambda_{34} \equiv \lambda_3 + \lambda_4$ and
$\lambda_{345} \equiv \lambda_3 + \lambda_4 + {\rm Re}(\lambda_5)$.
We should stress here that the relations~\eqref{eq:NAcond} must include or take place in a CP-invariant weak basis in which the relative CP-odd phase $\xi$ vanishes, i.e.~when $\xi =0$. Nevertheless, a covariant weak-basis independent condition for SM Higgs alignment can be formulated in terms of a vanishing commutator of two rank-2 Sp(4) tensors.

There are three distinct subgroups of NHAL that have been identified so far in the literature,
\begin{align}
  \label{eq:NHALs}
  \text{\bf Symmetry~8:} &\quad \text{Sp}(4)\, \simeq\, \text{SO}(5),\ \lambda_1 =\lambda_2 = \lambda_3/2\,,\nonumber\\
  \text{\bf Symmetry~4:} &\quad \text{SU(2)}_{\rm HF}\times \text{U}(1)_Y\, \simeq\, \text{O}(3)\times \text{O}(2)_Y,\ \lambda_1 = \lambda_2 = \lambda_{34}/2\,,\\
  \text{\bf Symmetry~2:} &\quad \text{CP}1\times \text{SO}(2)_{\rm HF}\times \text{U}(1)_Y\, \simeq\, \text{CP}1\times \text{O}(2)\times \text{O}(2)_Y,\
  \lambda_1 = \lambda_2 = \lambda_{345}/2\,. \nonumber
\end{align}
In~\eqref{eq:NHALs}, all $\text{SU}(2)_L$ gauge factors were suppressed, and their isomorphisms in the bilinear field space are given, as well as their listing in Table~\ref{tab:1} of all symmetries that can be relevant to NHAL. While the quartic coupling relations given in~\eqref{eq:NHALs} for Symmetries 8 and 4 remain invariant under $\text{SU}(2)$ reparameterisations of the doublets $\Phi_1$ and $\Phi_2$, this is no longer true for Symmetry 2, for which the relations~\eqref{eq:NAcond} modify in general. In order to put this on a more firm mathematical basis, let us denote the group of reparameterisations with $G_R$, where $G_R = \text{SU}(2)_{\rm HF}$ for the case at hand. Then, the parameter relations obtained from the action of a symmetry group $G$ do not alter, iff the criterion $ G_R\, \cap\, G\: =\: G_R$ is satisfied.
For Symmetry 2, this criterion is violated. In fact, SU(2) field transformations can also change the standard form of CP1. Consequently, any soft-breaking mass parameter introduced in the 2HDM potential must be real~\cite{Pilaftsis:2016erj}. But, as we will analyse in more detail in Section~\ref{sec:SO2HDM}, even in this case, SCPV typically leads to a VEV for $\Phi_2$ with $\xi \neq 0$, thereby triggering deviations from NHAL.

Symmetry groups~$G$ associated to Symmetries\,$3,5,6,7,8$ and their derivatives that emanate from different field-basis choices under the action of $G_R$ (see Table~\ref{tab:1}) are referred to as {\em custodial}~\cite{Pilaftsis:2011ed} since they involve generators that fail to commute with the generator~$T^0$ of the SM-hyper charge group: ${\text{U}(1)_Y \simeq\text{O}(2)_Y}$. These non-commuting generators can then be used to define the {\em coset space}: $G/\text{O}(2)_Y$, which does not form a {\em quotient group} in general. Hence, these generators are associated with transformations which are explicitly broken by the $\text{O}(2)_Y$ gauge coupling $g'$ that occurs in the gauge-kinetic terms of Higgs doublets. By contrast, Symmetries 4, 2 ($2'$, $2''$), and 1 ($1'$, $1''$) are O(2)$_Y$ invariant, and therefore include an  explicit O(2)$_Y$ factor as listed in Table~\ref{tab:1}. As a consequence of the above discussion, one must expect that NHAL arises from a custodial symmetry group that contains Symmetry~2 (i.e. CP1$\times$O(2)$\times$O(2)$_Y$).  Indeed, Table\ref{tab:1} confirms the existence of previously uncounted additional NHAL custodial symmetries
\begin{align}
  \label{eq:NHALnew}
\text{\bf Symmetry~7:} &\quad \text{CP1}\times \text{O}(4)\, \simeq\, \text{CP1}\times\text{Sp}(2)_{\Phi_1\Phi_2}\times \text{Sp}(2)_{\Phi_2\Phi_1}\;, \,\lambda_1 =\lambda_2 = \lambda_{3}/2\,,\ \lambda_4 = -\text{Re}(\lambda_5)\,.\nonumber\\
\text{\bf Symmetry~5:}&\quad \text{O}(2)\times \text{O}(3)\, \simeq\, \text{SO}(2)_{\rm HF}\times \text{Sp}(2)_{\Phi_1+\Phi_2}, \,\lambda_1 =\lambda_2 = \lambda_{345}/2\,,\ \lambda_4 = \text{Re}(\lambda_5) \;.\nonumber
\end{align}
We note the embedding: CP1$\times$O(2)$_Y\subset\text{Sp}(2)_{\Phi_1+\Phi_2}$. The latter as well as the custodial subgroups $\text{Sp}(2)_{\Phi_1\Phi_2},\,\text{Sp}(2)_{\Phi_2\Phi_1}  \subset \text{Sp}(4)$ are all defined in~\cite{Darvishi:2019dbh}.
Likewise, it is interesting to notice that Symmetry 2 goes to the higher Symmetry~5 in the limit: $\text{Re}(\lambda_5) \to \lambda_4$, and Symmetry 7 to 8, when $\lambda_4 \to 0$. 

We should clarify here that Symmetry 7 was classified before in~\cite{Pilaftsis:2011ed} as a subgroup of the symplectic group Sp(4), which is one of the three primary realizations for NHAL. Recently, a set of parameter relations similar to Symmetry 7 leading to NHAL was also observed in~\cite{Aiko:2020atr}, within 
the context of some twisted custodial group: $\text{SU}(2)_R\times \text{SU}(2)_L$ that includes the $\text{SU}(2)_L$ gauge group. In this respect, we should comment here that this twisted group is only a subgroup of the group: $\text{CP1}\times\text{Sp}(2)_{\Phi_1\Phi_2}\times \text{Sp}(2)_{\Phi_2\Phi_1}\times \text{SU}(2)_L$,
as~specified in~\eqref{eq:NHALnew}.
Moreover, to the best of our knowledge, Symmetry~5 given in~\eqref{eq:NHALnew}, which was tabulated in~\cite{Battye:2011jj,Pilaftsis:2011ed}, has not been analysed before in connection with NHAL. 

\begin{table}[t]
\small
\begin{center}
\begin{tabular}{ |c|| l | l | l | l | } 
 \hline
No. & Syms with NHAL & No. & Syms breaking of NHAL & Types of CPV after SB\\ 
\hline \hline
8 & SO(5)& \multicolumn{3}{c|}{---} \\
\hline
7 &  CP1$\times$O(4) & 6 & SO(4)& ECPV(MCPV)
\\
7$'$ &  CP2$\times$O(4)$'$  & 6$'$ & SO(4)$'$& SCPV\\
\hline
5 &   O(2$)\times$O(3)  & 6$'$ & SO(4)$'$& SCPV
\\
5$'$  & O(2$)'\times$O(3) & 6 & SO(4)& ECPV(MCPV)\\
\hline
4 &  O(3)$\times$O(2)$_Y$ & \multicolumn{3}{c|}{---} \\
\hline
2 &  CP1$\times$O(2)$\times$O(2)$_Y$ & 1 & O(2)$\times$O(2)$_Y$ & ECPV(MCPV)\\
2$'$  &  CP2$\times$O(2)$'\times$O(2)$_Y$  & 1$'$  & O(2)$'\times$O(2)$_Y$ & SCPV\\
\hline
\end{tabular}
\end{center}
\caption{\textit{The continuous symmetries resulting from the CP breaking of NHAL symmetries are presented. The final column indicates the possible types of CPV realised after the introduction of the soft symmetry-breaking mass terms.}}
\label{tab:2}
\end{table}

In Table~\ref{tab:2}, we have explored whether the five NHAL symmetries given in~\eqref{eq:NHALs} and~\eqref{eq:NHALnew} (and those descending from different field-basis choices) can lead to physical~CPV or not. However, we must emphasise that not all NHAL symmetries can lead to CPV, after introducing soft symmetry-breaking masses, but only those associated with CP1 and CP2 symmetries. For instance, Symmetries 8 and 4 cannot source CPV, unless an explicit hard-breaking of symmetries is considered. Instead, as shown in the fourth column of Table~\ref{tab:2}, Symmetries $2$ and $2'$ can
break to the lower symmetries $1$ and $1'$, after specific operators (breaking softly or explicitly $2$ and $2'$) are added to the potential. Interestingly enough, Symmetries~$6,\,6'$ will transition to the lower symmetries~$5',\,5$, when the symmetry-breaking parameters of the latter are switched off from the potential. This paradox can be resolved by noting that all $6$-type symmetries are based on $\text{SO}(4) \simeq \text{SO}(3)\times \text{SO}(3)$ containing two custodial factors that produce no new restrictions on a U(1)$_Y$-invariant 2HDM potential. They could only produce non-trivial constraints on the new theoretical parameters present in a hypothetical U(1)$_Y$-violating 2HDM potential${}$~\cite{Battye:2011jj}.  

The types of CPV that can be realised by adding now all possible soft symmetry-breaking mass terms have been catalogued in the last column of Table~\ref{tab:2}. Our approach to maximising CPV should be characterised as natural according to 't Hooft's naturalness criterion since an enhanced symmetry gets realised once all CPV terms are switched off.

\setcounter{equation}{0}
\section{\boldmath O(2)$\times$O(2)$_Y$ and SO(4) Symmetric 2HDMs}
\label{sec:SO2HDM}
In this section, we focus on approximate symmetric 2HDM scenarios that allow for observable CP violation (CPV), as indicated in the fourth column of Table~\ref{tab:2}. Our main goal is to determine whether there exists a region of parameter space that can accommodate sizable CPV, while remaining consistent with the non-observation of an electron EDM, the SM alignment limit, and other LHC constraints. In the previous section, we identified two new custodial NHAL symmetries (in addition to the previously discussed CP1$\times$SO(2)$\times$O(2)$_Y$) based on the product groups CP1$\times$O(4) and O(2)$\times$O(3) [cf.\eqref{eq:NHALnew}]. From Table~\ref{tab:2}, one sees that CP1$\times$SO(2) and CP1$\times$O(4) break into the groups O(2)$\times$O(2)$_Y$ and SO(4), respectively. As will be demonstrated below, these two groups enable one to build minimal scenarios of soft and explicit CP breaking while maximising CPV in the scalar potential.

Let us first consider the parameter space of an O(2)$\times$O(2)$Y$-symmetric 2HDM. In the bilinear field formalism, this model is invariant under transformations generated by $T^0$ and $T^1$, which correspond to 2D rotations in the $R^0$-$R^2$ plane. In the original scalar field space, this is equivalent to 
\begin{align}
\Phi_\pm\, =\, \,\big(\Phi_1 \pm i \Phi_2\big)/\sqrt{2}\ &\to\ 
\Phi'_\pm\, =\, e^{\pm i \alpha}\, \big(\Phi_1 \pm  i\Phi_2 \big)/\sqrt{2}\,.
\end{align}
Therefore, in an O(2)$\times$O(2)$_Y$-invariant 2HDM, 
the potential parameters obey the following relations:
\begin{eqnarray}
{\rm O}(2)\times {\rm O}(2)_Y:  \quad m_{11}^2 = m_{22}^2,\ \ {\rm Im}(m_{12}^{2}),\  \ \lambda_1 = \lambda_2 = \lambda_{345}/2,\ \ {\rm Im}(\lambda_{6}) = {\rm Im}(\lambda_{7})\,.
\end{eqnarray}
Although the naively CP-odd parameter $\mathrm{Im}(m_{12}^2)$ remains independent here, the O(2)$\times$O(2)$Y$-symmetric potential itself is CP-conserving. One can break this symmetry softly by introducing non-zero mass terms, $\mathrm{Re}(m_{12}^2)$ and $m_{11}^2 - m_{22}^2 \neq 0$, thereby inducing a non-removable CP-odd phase in the 2HDM potential.

We now turn to the SO(4)-symmetric 2HDM. The parameter space of this model is constrained by the action of the $T^{0,2,5,7,8,9}$ generators, corresponding to Symmetry~6 in Table\ref{tab:1}. Consequently, the potential parameters of an SO(4)-symmetric 2HDM must satisfy 
\begin{equation}
\mathrm{SO}(4):\quad m_{11}^2 = m_{22}^2, \,\,  \mathrm{Im}(m_{12}^{2}), \,\, \lambda_1 = \lambda_2 = \lambda_{3}/2,\,\, \mathrm{Re}(\lambda_{5}) = -\lambda_4,\,\,  \mathrm{Im}(\lambda_{6}) = \mathrm{Im}(\lambda_{7})\,.
\end{equation}
As before, one may introduce soft symmetry-breaking mass terms in the $\mathrm{SO}(4)$-symmetric potential to allow for CPV. In this scenario, the breaking pattern $\mathrm{SO}(4) \to \mathrm{O}(2)\times \mathrm{O}(3) \to \mathrm{SO}(3)$ induces small departures from NHAL, while simultaneously violating the CP symmetry of the 2HDM potential. By choosing a weak basis where $\lambda_6 = \lambda_7$, it is then possible to realise all three types of CP violation.

\begin{table}[t!]
\centering
\resizebox{1\textwidth}{!}{
\begin{tabular}{|c|c|c|c|c|c|c|c|c|c|c|}
\hline
Symmetries & CPV  & $\tan\beta$ & $\xi$ & $\phi_{12}$ & $M_{H_1}$ & $M_{H_2}$ & $M_{H_3}$ & $M_{H^\pm}$ & $g_{H_1VV}$ & EDM [$e\cdot {\rm cm}$] \\
\hline \hline
\multirow{2}{*}{O(2)$\times$O(2)$_Y$} & ECPV & 1.608 & 0.002 & -0.05  & 125.10 & 682.99 & 787.67 & 640.18 & 0.998 & $9.51 \times 10^{-30}$ \\
& MCPV & 0.804 & -1.93& 0.27 & 125.21 & 225.11 & 270.50 & 380.54 & 0.989 & $7.67 \times 10^{-30}$  \\
O(2)$'\times$O(2)$_Y$& SCPV & 0.71 & 0.67 & 0  & 125.52 & 275.23 & 448.01 & 551.38 & 0.957 & $3.86 \times 10^{-30}$  \\
\hline
\multirow{2}{*}{SO(4)} & ECPV & 0.827 & 0.002 & -0.02 & 125.28 & 506.93 & 570.58 & 570.59 & 0.995 & $2.87 \times 10^{-30}$  \\
& MCPV & 0.740 & -1.25 & 0.44  & 125.40 & 408.49 & 433.98 & 433.99 & 0.984 & $1.07 \times 10^{-30}$  \\
SO(4)$'$& SCPV & 0.742 & -2.42 & 0 & 125.68 & 216.63 & 265 & 442.21 & 0.948 & $3.55 \times 10^{-30}$  \\
\hline
\end{tabular}}
 \caption{\textit{Benchmark scenarios used in Figures~\ref{ECPV-AL}, \ref{MCPV-AL} and \ref{SCPV-AL}. All angles are in radians and all masses are in GeV.}}
\label{tab:BMs}
\end{table}

To assess the allowed size of CPV, we will scan the model parameters and ensure that they align with experimental constraints on gauge couplings.
The key model parameters for our scan are
$(\tan\beta\,,\ M_H^\pm\, ,\ \lambda_{3}\, ,\ \lambda_{4}\,,\ \xi\,,\ \phi_{12})$ with the following restrictions on the parameters:
\begin{equation}
M_H^\pm \in [250, 800]\, {\rm GeV}\,,\quad \beta \in \left(0, \pi/2\right), \quad \{\xi,\phi_{12}\} \in (-\pi, \pi]\;,
\end{equation}
along with $M_{H_1} = 125.46 \pm 0.35$ GeV. As for the quartic couplings $\lambda_{3,4}$, these are only constrained to lie in the perturbative regime, i.e.~$\lambda_{3,4} < 4\pi$. For the SO(4)-symmetric scenarios, the relationship $\lambda_{3} = 2\lambda_{1}$ holds. Consequently, after a careful parameter scan, the benchmark models are summarised in Table~\ref{tab:BMs}.

In addition, the stringent 90\% confidence-level limit on the electron EDM~\cite{Andreev:2018ayy}, \begin{align} \label{eq:exp-edm} |d_e| < 1.1 \times 10^{-29}\ e \cdot \text{cm}, \end{align} imposes strong constraints on any new source of CPV, including those originating from the 2HDM potential. Within the SM, the electron EDM lies many orders of magnitude below the current experimental sensitivity. Consequently, observing a non-zero electron EDM would be a clear signal of new physics, potentially arising from an extended CPV scalar sector of the 2HDM. In Type-II 2HDMs, the dominant contribution to the electron EDM arises at two loops via the Barr--Zee mechanism~\cite{Barr:1990vd,Pilaftsis:1999qt,Abe:2013qla,Darvishi:2022wnd}, whereas one-loop contributions to $d_e$ are comparatively suppressed. Nevertheless, with certain fine tuning of the parameter space, one-loop contributions to the electron EDM can dominate in a generic Yukawa interaction of 2HDM~\cite{Darvishi:2022wnd}. Analytic expressions for the two-loop contributions are given in Ref.~\cite{Darvishi:2023fjh}.

\begin{figure}[t]
\centering
\includegraphics[width=0.85\textwidth]{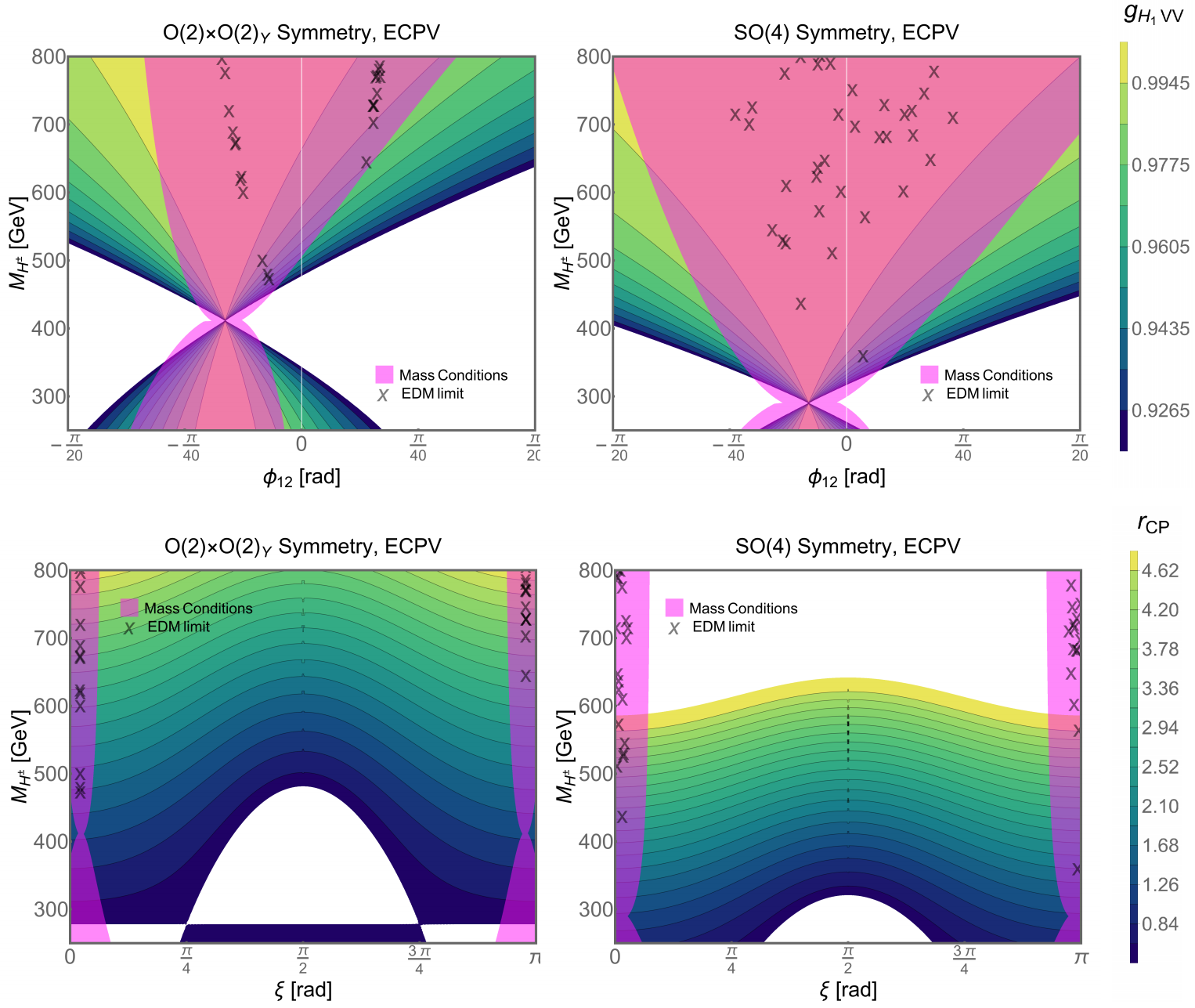}
\caption{\textit{The gauge boson coupling $g_{H_{1}VV}$ and $r_{\rm CP}$ in the ($M_H^\pm, \xi$)-plane are displayed for different types of ECPV in O(2$)\times$O(2)$_Y$- and SO(4)-softly broken cases. The cross symbols ``$\times$'' showcase benchmark points consistent with the EDM limit of Electron in ~\eqref{eq:exp-edm}. }}
\label{ECPV-AL} 
\end{figure}

{\allowdisplaybreaks
\begin{figure}[t]
\centering
\includegraphics[width=0.85\textwidth]{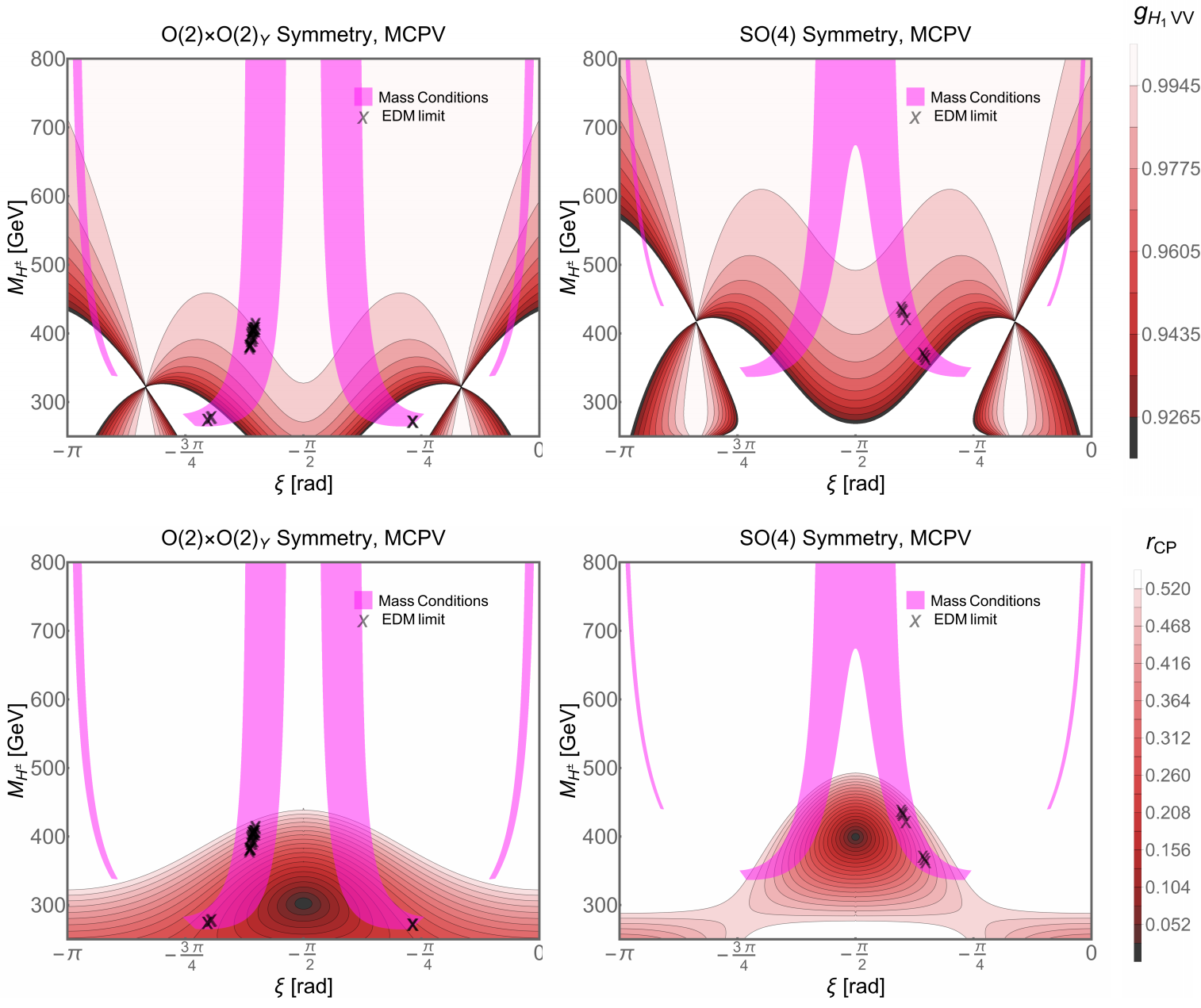}
\caption{\textit{The same as in Figure~\ref{ECPV-AL}, but for MCPV.}}
\label{MCPV-AL} 
\end{figure}}

{\allowdisplaybreaks
\begin{figure}[t]
\centering
\includegraphics[width=0.85\textwidth]{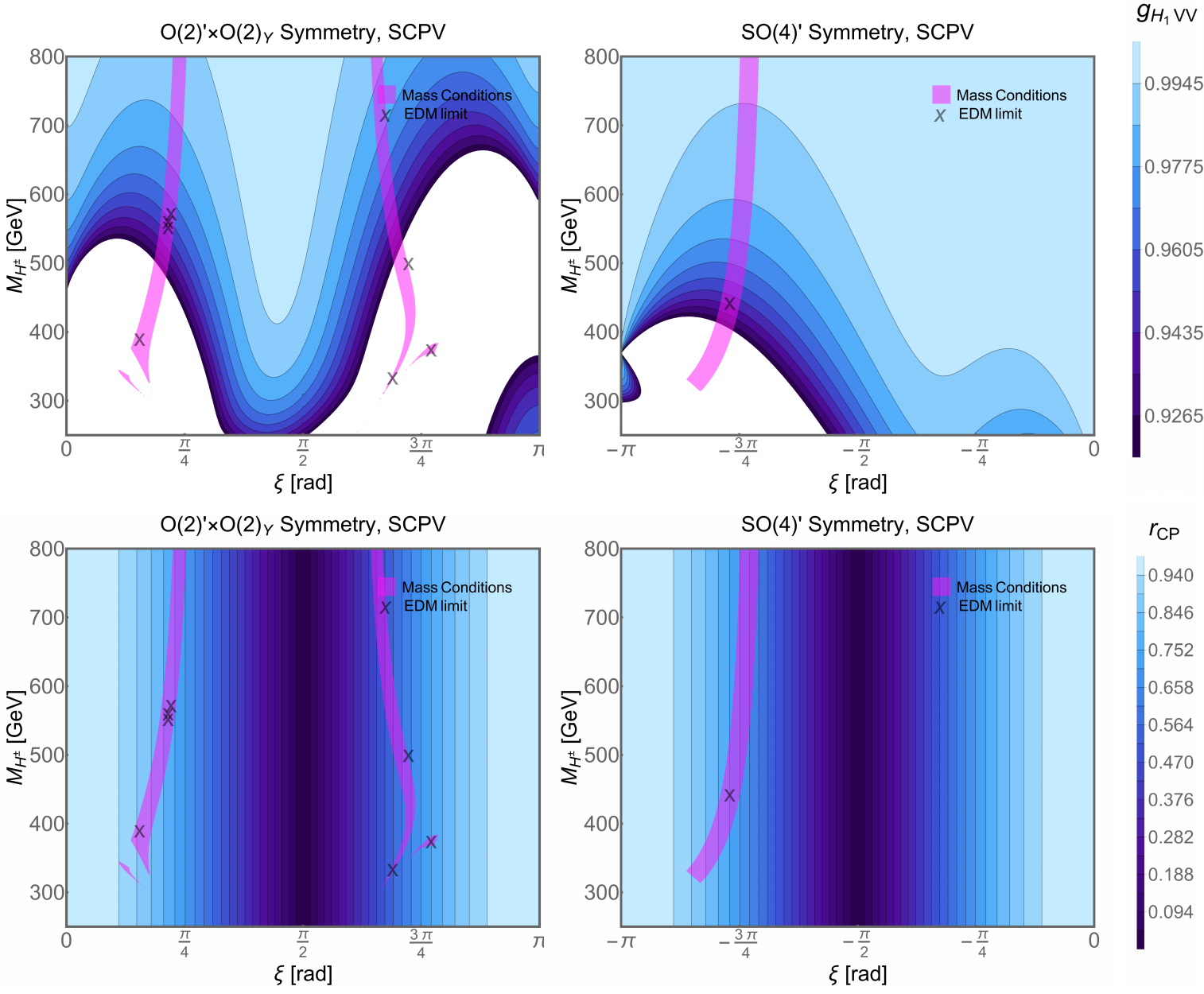}
\caption{\textit{The same as in Figure~\ref{ECPV-AL}, but for SCPV in O(2)$'\times$O(2)$_Y$- and SO(4)$'$-softly broken cases.}}
\label{SCPV-AL} 
\end{figure}}

In Figures \ref{ECPV-AL}, \ref{MCPV-AL}, and \ref{SCPV-AL}, the upper panels present the predicted values of the SM-normalised coupling $g_{H_1VV}$ for the lightest $H_1$ boson to the gauge bosons $V = W^\pm, Z$, while the lower panels show the parameter $|r_{\rm CP}|$. These results are shown for the O(2)$\times$O(2)$_Y$ (left panels) and SO(4) (right panels) symmetric 2HDMs, considering three different types of CP-violating scenarios (ECPV, MCPV, and SCPV) depicted in green, red, and blue, respectively. In our numerical analysis, the CP-odd phases $\phi_{12}$ and $\xi$ are varied to satisfy the two mass conditions $M_{H_1} \approx 125$ GeV and $M_{H_3} > M_{H_2}$. The plots also display a set of benchmark points marked by cross symbols ``$\times$'', which are compatible with the electron EDM limit in Eq.\eqref{eq:exp-edm}. It should be noted that the existence of these benchmark points does not imply that the rest of the parameter space is necessarily validated or excluded. In a 2HDM scenario with ECPV, the phase $\phi_{12}$ is pivotal, as illustrated in the upper panels of Figure \ref{ECPV-AL}, where the SM-normalised $H_1VV$ coupling $g_{H_1VV}$ is projected onto the $(\phi_{12},M_{H^\pm})$-plane. By analogy, Figures \ref{MCPV-AL} and \ref{SCPV-AL} provide corresponding projections for the MCPV and SCPV scenarios, respectively, showing how $g_{H_1VV}$ is distributed on the $(\xi,M_{H^\pm})$-plane. In these plots, the violet areas highlight viable regions that satisfy $M_{H_1} = 125.46 \pm 0.35$~GeV and $M_{H_3} > M_{H_2} > 200$~GeV, while uncolored regions are excluded from the analysis. Figures \ref{ECPV-AL}, \ref{MCPV-AL}, and \ref{SCPV-AL} show that certain benchmark points can satisfy all the relevant constraints at the same time. The lower panels of these figures provide more detail, showing the value of $|r_{\rm CP}|$ for each point in the top panels. In particular, the white area in the bottom-right panel of Figure~\ref{ECPV-AL} indicates where $|r_{\rm CP}| > 5$, as discussed in Section~\ref{sec:CPV2HDM}. Scenarios with $\lambda_4 = -{\rm Re}(\lambda_5)$ display a mass degeneracy between the charged and heavy neutral Higgs bosons, which is lifted once the phase~$\xi$ departs significantly from zero. The relevant benchmark scenarios under consideration are listed in Table~\ref{tab:BMs}, including an SO(4)$'$-symmetric 2HDM where $\lambda_4 = {\rm Re}(\lambda_5)$, as indicated in Table~\ref{tab:1}.

\section{Conclusions}\label{sec:Concl}

We have presented a detailed study of the Two Higgs Doublet 
Model~(2HDM), placing special emphasis on the vacuum topology of its potential and its implications for CP violation (CPV). 
We have rigorously identified that in general three types of CPV exist: (i) Spontaneous CPV, (ii) Explicit CPV and (iii) Mixed Spontaneous and Explicit CPV (MCPV). In all these three different scenarios, only two CPV phases can remain independent.

The vacuum topology of a general 2HDM potential is usually set by the action of some accidental symmetry, as well as the size of any explicit or spontaneous breaking parameters, such as the CP-odd phases $\xi$ and $\phi_{12}$. A key result of this work has been to illustrate the pivotal role of the complex parameter $r_{\rm CP}$ (defined in~\eqref{eq:rCP}), which enables us to reliably distinguish the three types of CPV mentioned above. In particular, both the magnitude $|r_{\rm CP}|$ and its CP-odd phase $\phi_{\rm CP}$ serve as faithful indicators of the CPV realization in the 2HDM. Although it is well known that $|r_{\rm CP}|>1$ allows only an ECPV scenario, the region $|r_{\rm CP}|<1$ can exhibit much richer structures if explicit CPV sources are present. Specifically, for $1/2<|r_{\rm CP}|<1$, any of the SCPV, ECPV, or MCPV scenarios may occur, depending on the value of $\phi_{\rm CP}$. In contrast, if $0<|r_{\rm CP}|<1/2$, only SCPV and MCPV remain possible, and there is always a component of SCPV in this case. Finally, if $r_{\rm CP}$ is tuned to zero, the potential's topology becomes more intricate, so higher-order corrections beyond the Born approximation are necessary.

An important phenomenological constraint on any theory of new physics is the required\- alignment of the coupling strengths of the observed 125-GeV scalar resonance to the $W^\pm$ and $Z$ gauge bosons with strengths as predicted by the SM. In the 2HDM, such an SM-Higgs alignment can be naturally realised through certain accidental symmetries~\cite{BhupalDev:2014bir}. However, their imposition unavoidably leads to a CP-conserving 2HDM. Here, we have revisited these NHAL symmetries in the bilinear covariant formalism to explore how to maximise CPV via minimal soft and explicit breakings to lower symmetries. In particular, we recovered the two known NHAL symmetries, 
i.e.~Sp(4) and SU(2)$_{\rm HF}$ in the original field basis. But when CP is imposed on a general 2HDM potential thanks to a CP1 or CP2 symmetry, we have then found that besides CP1$\times$O(2)$\times$O(2)$_Y$, there exist two new custodial NHAL symmetries that make use of the product groups: $\text{CP1}\times\text{O}(4)$ and $\text{O}(2)\times \text{O}(3)$ [cf.~\eqref{eq:NHALnew}]. In fact, the groups, CP1$\times$O(2)$\times$O(2)$_Y$ and CP1$\times$O(4), break into the key symmetry groups, O(2)$\times$O(2$)_Y$ and SO(4), respectively. These smaller groups provide minimal frameworks for introducing soft and explicit CP breakings, thereby maximising CPV in the scalar potential.

We have derived upper bounds on key CPV parameters that reflect SM misalignment, from the non-observation of an electron EDM in representative 2HDM scenarios based on the two aforementioned symmetry groups. In addition, we have delineated the corresponding CPV parameter space of such approximate NHAL scenarios, which can be probed at the LHC.

\subsection*{Acknowledgements} 
\noindent
STFC supports ND and AP work under grant numbers ST/Y004590/$1$ and ST/X00077X/$1$, respectively.

\bibliographystyle{unsrt}
\bibliography{biblio-NT}

\end{document}